\documentstyle[12pt]{article}

\def\endpage{\vfill\eject}

\begin{document}
\rightline {published in Phys. Lett. {\bf B354} (1995) 111-116}
\vskip 2. truecm
\centerline{\bf Spontaneous Symmetry Breaking in Fermion-Gauge Systems:}
\centerline{\bf A Non Standard Approach}
\vskip 2 truecm

\centerline{ Vicente Azcoiti $^a$, Victor Laliena $^a$
and Xiang-Qian Luo $^b$}

\vskip 0.15 truecm
\centerline {\it  $^a$Departamento de F\'\i sica Te\'orica, Facultad de
Ciencias, Universidad de Zaragoza,}
\centerline {\it 50009 Zaragoza, Spain}
\vskip 0.15 truecm
\centerline {\it  $^b$HLRZ, c/o KFA, D-52425 J\"ulich, Germany}
\vskip 3 truecm
\centerline {ABSTRACT}
We propose a new method for the study of the chiral
properties of the ground state in \rm{QFT's} based on the computation
of the probability distribution function of the chiral condensate.
It can be applied directly in the chiral limit and
therefore no mass extrapolations are needed. Furthermore this approach allows
to write up equations relating the chiral condensate with quantities
computable by standard numerical methods, the functional form of these
relations depending on the broken symmetry group. As a check, we report some
results for the compact Schwinger model.

\vfill\eject

\par
The study of the chiral properties of the ground state in
Quantum Field Theories (\rm{QFT's})
is one of the main points of attention in recent
developments of Lattice Gauge Theories ($LGT's$).
In $QCD$, the gauge theory for the
strong interacting sector of the Standard Model, the low energy physics
could not be understood without a detailed analysis of the chiral properties
of the vacuum. Furthermore there are strong evidences from numerical
simulations of $QCD$ at finite temperature, suggesting that a chiral transition
appears at some critical temperature $T_c$ separating the quark-gluon plasma
phase from the hadronic phase.

Concerning the abelian model, recent numerical studies in the noncompact
formulation \cite{NCQED4} show the existence of a strongly coupled
$QED$ and open
the possibility to get a non-trivial continuum limit for a non asymptotically
free gauge theory in four dimensions. The strongly coupled phase of this
model is characterized by a vacuum state not invariant under chiral
transformations and therefore a mechanism for dynamical mass generation
appears.

Last, the analysis of the chiral properties of the vacuum in $2+1$
\rm{QFT's} and
of their dependence on the flavor number has also become of great interest
because of its possible relevance in the high $T_c$ superconductivity
phenomenon.

In this paper we propose a new method for the study of the chiral
properties of the ground state in $QFT's$ which is based on the computation
of the probability distribution function $(p.d.f.)$ of the chiral condensate
in the chiral limit. This is an standard procedure when analysing spontaneous
symmetry breaking in spin systems or in $QFT's$ with bosonic degrees of
freedom, since this kind of degrees of freedom can be simulated directly in
a computer. However, in in the numerical simulations of a $LGT$ with
dynamical fermions, the Grassmann fields must be integrated analytically for
obvious reasons. Then, even if the ground state is degenerate, in the
analytical procedure we integrate over all possible vacuum states,
the chiral order parameter being always zero for massless fermions. We will
show here how despite of the fact that Grassmann variables
cannot be simulated in a
computer, an analysis of spontaneous symmetry breaking without a symmetry
breaking external field, which is standard in te case of spin systems,
can also be done in $QFT's$
with fermion degrees of freedom. Previous attemps to analyse the vacuum
structure in fermion-gauge systems \cite{SMIT} were based on several
approximations. We want to remark that our analysis is completely free from
approximations.

The main advantage of this method, when compared with
standard simulations, is that we can work directly in the chiral limit and
therefore no mass extrapolations are needed. Furthermore this approach allows
to write up equations relating the chiral condensate with quantities
computable by standard numerical methods, the functional form of these
relations depending on the nature of the broken symmetry group.

Another interesting field of application of $p.d.f.$ of order parameters is
the quantitative description of interacting social groups in
stochastic models for the formation of Public Opinion. Previous work in
this field \cite{PO} analyses the time dependent $p.d.f.$ of public opinion
for the simplest case of two kind of opinions $(+,-)$, modeling it with an
Ising ferromagnet, the dynamics of which takes into account the influence on
an individual of its neighbors. Generalization of this work to the case of
a continuous opinion (continuous symmetry) could be another possible
application of the formalism developed here. The application of fermionic
models to Public Opinion formation is interesting not only for the continuous
character of the symmetry but also because of the long range interactions
induced by the fermion dynamics, which would give a more precise
description of the high degree of long distance communication in present
world.

\vskip 1truecm
\leftline{\bf 1. Theoretical Grounds}
\par
Our starting point is a $QFT$ describing a gauge field coupled to a fermion
matter field and regularized by means of a space-time lattice. Next suppose
the ground or equilibrium state of this model is degenerate, and be $\alpha$
the index which characterize all possible vacuum states. The ensemble of all
equilibrium states is a partition of the Gibbs state and the probability
$w_{\alpha}$ to get the vacuum state $\alpha$ when choosing randomly an
equilibrium state depends on the total free energy of the $\alpha$ state
\cite{PARISI}.

Since we are interested in the analysis of the chiral properties of the
ground state, let us choose as order parameter the chiral condensate
$\bar{\psi}\psi$ and characterize each vacuum state $\alpha$ by the
expectation value $c_{\alpha}$ of the order parameter in the $\alpha$ state,
i.e.

\begin{eqnarray}
c_{\alpha} = {1\over N} \sum_{x} <\bar{\psi}(x)\psi(x)>_{\alpha}
\label{1}
\end{eqnarray}

\noindent
where $N$ is the total number of lattice sites and the sum is over all
lattice points. The $p.d.f.$ $P(c)$ of the chiral order parameter $c$ will
be given by

\begin{eqnarray}
P(c) = \sum_{\alpha} w_{\alpha}\delta (c - c_{\alpha}).
\label{2}
\end{eqnarray}

The function $P(c)$ tells us what is the probability that choosing randomly
a vacuum state, we get the value $c$ for the chiral order parameter. If the
vacuum state is invariant under chiral transformations, i.e., if it is
unique as concerning the chiral symmetry, $P(c)$ will be a single $\delta$
function $\delta (c)$. Otherwise $P(c)$ will be a more complex function,
sum of $\delta$ functions in the case of a discrete symmetry group or a
continuous function in the other cases.

The next step now is to relate $P(c)$ with magnitudes which can be computed
by numerical simulations. To this end, let us write

\begin{eqnarray}
P(c) = {lim_{N\rightarrow\infty}}< \delta ({1\over N} \sum_{x}
\bar{\psi}(x)\psi(x) - c)>
\label{3}
\end{eqnarray}

\noindent
where the expectation value in (\ref{3})
is computed in the Gibbs state and the
integration measure is that associated to the partition function

\begin{eqnarray}
{\cal Z} = \int d{\bar\psi} d{\psi} dU e^{-S_{G}(U) + \bar\psi\Delta\psi}.
\label{4}
\end{eqnarray}

\noindent
$S_G$ in (\ref{4})
is the pure gauge action and $\Delta$ the fermionic matrix.

To check that equation (\ref{3})
 gives us indeed the $p.d.f.$ of the chiral condensate, it
is enough to verify that all the moments of the $p.d.f.$ defined in
(\ref{2})
agree with those of (\ref{3}).
The verification for the first moment is trivial
whereas for the higher moments it is enough to take into account that in the
thermodynamical limit intensive quantities do not fluctuate in  any
equilibrium state $\alpha$, i.e.

\begin{eqnarray*}
{1\over{N^p}}< \sum_{x_1,x_2,...x_p} \bar{\psi}(x_1)\psi(x_1)
\bar{\psi}(x_2)\psi(x_2).....\bar{\psi}(x_p)\psi(x_p)>_{\alpha}
\end{eqnarray*}
\begin{eqnarray}
={1\over{N^p}} \sum_{x_1,x_2,...x_p} <\bar{\psi}(x_1)\psi(x_1)>_{\alpha}
<\bar{\psi}(x_2)\psi(x_2)>_{\alpha}.....<\bar{\psi}(x_p)\psi(x_p)>_{\alpha}.
\label{5}
\end{eqnarray}

Expression (\ref{3}) is not suitable for numerical computation. However we can
define its Fourier transformed $P(q)$

\begin{eqnarray}
P(q) = {\int e^{iqc} P(c) dc}
\label{7}
\end{eqnarray}

\noindent
which, as will be shown, can be numerically computed.

Since we are interested in the analysis of the chiral symmetry on the lattice,
we will use the staggered fermions regularization. The fermion matrix $\Delta$
can be written as

\begin{eqnarray}
\Delta = m + i\Lambda
\label{8}
\end{eqnarray}

\noindent
where $m$ is the fermion mass and $\Lambda$ a hermitian matrix which depends
on the gauge field configuration. The eigenvalues of $\Lambda$ are real and
symmetric. Taking into account all these properties of $\Delta$, the
following expression for $P(q)$ can be derived from (\ref{7})

\begin{eqnarray}
P(q) = <\prod_{j} (1 - {{q^{2} - i2mqN} \over
{N^{2} (m^{2}+\lambda^{2}_{j})}})>
\label{9}
\end{eqnarray}

\noindent
and in the chiral limit

\begin{eqnarray}
P(q) = <\prod_{j} (1 - {{q^{2} } \over
{N^{2} \lambda^{2}_{j}}})>
\label{10}
\end{eqnarray}

\noindent
where the product in (\ref{9}) and (\ref{10})
runs over all positive eigenvalues $\lambda_j$
and the mean values are computed with the probability distribution function
of the effective gauge theory obtained after integrating out the fermion
fields. In order to get meaningful results for (\ref{10}), special care must be
taken with the boundary conditions for the fermion field to
avoid zero modes.

The function $P(q)$ can be computed numerically and then, by inverse
Fourier transform we get $P(c)$.
We can go deeper in the investigation of the form of the $p.d.f.$ $P(c)$. In
the lattice regularized action of a gauge theory with Kogut-Susskind fermions,
only a $U(1)$ subgroup of the continuous chiral symmetry is preserved. Then
if this continuous residual symmetry is spontaneously broken, we will get
a continuum of equilibrium states characterized by an angle $\alpha$ with
$\alpha$ taking values between $-\pi$ and $\pi$. The $v.e.v.$
of the chiral condensate at each vacuum will be given by
$c_0 \cos (2\alpha)$, $c_0$ being the value corresponding to the
$\alpha-vacuum$ selected when switching-on an external {\it ``magnetic"}
field. Therefore, the function $P(c)$ can be computed as

\begin{eqnarray}
P(c) = {1\over{2\pi}} \int^{\pi}_{-\pi} {d\alpha
\delta(c-c_0\cos(2\alpha))}
\label{11}
\end{eqnarray}

\noindent
which gives for $P(c)$ the value $1/(\pi(c^{2}_0-c^{2})^{1/2})$ for
$-c_{0}\leq c \leq c_{0}$, $P(c) = 0$ otherwise (see Fig. 1).
Its Fourier transformed

\begin{eqnarray}
P(q) = {1\over{2\pi}} \int^{\pi}_{-\pi} {d\theta e^{iqc_0\cos\theta}}
\label{12}
\end{eqnarray}

\noindent
is the well known  zeroth order Bessel function of the first kind $J_0(qc_0)$.

Several relations between the chiral condensate and the eigenvalues of the
fermionic matrix can be derived from (\ref{12}).
For example, by doing the second
derivative of the function $P(q)$ we get the second moment of the
distribution $P(c)$ and then the following relation holds

\begin{eqnarray}
{c_0}^2 =
{<\bar{\psi}\psi>}^2 = {<{4\over{N^{2}}}\sum_{j} {1\over{\lambda^{2}_{j}}}>}.
\label{13}
\end{eqnarray}

\noindent
In the symmetric phase, the right hand side of
(\ref{13}) is the longitudinal susceptibility
normalized by the lattice volume $N$.
Furthermore the function $P(q)$ is proportional to the partition function
$\cal Z$ evaluated in the imaginary axis of the complex mass plane and
therefore a relation between the zeroes of the partition function in the
complex mass plane and the value of the chiral condensate follows from this
relation.

Using (\ref{12}) the following relations between the chiral condensate
$c_0$, the zeroes $q_i$ of $P(q)$ and the zeroes $a_i$ of the zeroth order
Bessel function of first kind can be derived,

\begin{eqnarray}
c_{0}(i) = {a_{i}\over{q_{i}}}.
\label{14}
\end{eqnarray}

\noindent
In the thermodynamical limit the value of $c_0$ will be independent of the
order $i$ of the zero in (\ref{14}). At finite volume $c_0$ depends on $i$.
However we have
observed that a plateau appears in the plot of $c_0$ as a function of $i$ when
chiral symmetry is spontaneously broken. The extent of this plateau increases
with the lattice size becoming eventually infinite for infinite lattice
volumes. These things are very well illustrated in Fig. 2 where we plot some
preliminary results on the lattice
chiral condensate $c_0(i)$ against $i$ for the one-flavor compact Schwinger
model, obtained from $MFA$ simulations \cite{NOS}.
The continuous line in this figure stands for the
continuum analytical result (in units of charge) times $\beta^{-1/2}$.
Of course the Schwinger model is not very interesting from a physical point
of view but, as well known, it is a very good laboratory to check new
proposals for two reasons: exact analytical results are available and it
shares many interesting properties with other more relevant physical models.
This is the reason why we believe the results reported in Fig. 2 are very
encouraging.

\vskip 1truecm
\leftline{\bf 2. Discrete versus continuous p.d.f.}
\par
The analysis of $p.d.f.$ of order parameters in Statistical Mechanics is
not new. It has been developed for ferromagnetic Ising-like systems
\cite{BINDER} and also for spin-glasses \cite{PARISI}, the last with a much
more complicated vacuum structure than magnets. However this is the first
time to our knowledge this kind of analysis is applied, without any
approximation, to $QFT's$ with dynamical fermion fields.

There are several important differences between $p.d.f.'s$ in spin systems
and fermionic systems and we would like to point out them here. First and
from a practical point of view, the $p.d.f.$ of the density of magnetization
in a spin system can be directly measured by computer simulations whereas
in a fermionic system this is not possible since Grassmann variables cannot
be simulated in a computer. This is the reason why in our case we
work with the Fourier transform $P(q)$ (\ref{10}) rather
than with $P(c)$ (\ref{3}).
Furthermore and as a consequence of the anticommuting character of fermionic
fields, all the moments of the $p.d.f.$ $P(c)$ of order larger than the
lattice volume $N$ vanish since they are correlation functions of
anticommuting variables with some repeated index. In fact the function
$P(q)$ given by equation (\ref{10}) is a polynomial of degree
the lattice volume
$N$ and therefore its inverse Fourier transform does not exists as a
regular function if $P(q)$ is defined in an unbounded momentum space.

This peculiarity of the $p.d.f.$ for fermionic systems complicates the
finite size scaling analysis of it since in order to do such a kind of
analysis, we need to put an arbitrary cut-off in momentum space which
could induce extra finite-volume effects. However we can by-pass
this difficulty by applying the finite size scaling hypothesis, instead
to the $p.d.f.$ $p(c)$ as in Binder's work on Ising systems \cite{BINDER},
to its Fourier transform $P(q)$. Then the finite size scaling hypothesis
for $P(q)$ reads

\begin{eqnarray}
P_{L}(q) = {f(qL^{-\beta\over{\nu}},{\xi\over{L}})}.
\label{15}
\end{eqnarray}

\noindent
where $L$ is the linear dimension of the lattice and $f$
is a universal scaling function. Equation (\ref{15}) should work
for $L$ large and near the critical point.

The next step now should be to estimate the explicit form of the universal
function $f$ in both the symmetric and broken phases. Since this
kind of analysis is too long to be reported here, we will stress only some
important differences which appear when comparing with Binder's analysis
of Ising like systems. The physical origin of the different behavior of
the universal function $f$ for Ising and fermion-gauge systems is
the existence of a continuum of equilibrium states in the broken phase
of fermion-gauge systems in contrast to the two ground states of the low
temperature phase of the Ising model.

The symmetric phase of both models is characterized by a non degenerate
vacuum and Binder's analysis \cite{BINDER}, which is based on the assumption
that in this case $P_L(c)$ (equivalently $P_L(q)$) is well approximated by
a gaussian function centered at the origin, should work also for fermionic
systems, at least for not too large values of $q$. The situation however
changes drastically in the broken phase where Binder's analysis, based on
the two gaussian approximation of the $p.d.f.$ at finite lattice size, fails
completely to describe the spontaneous breaking of a continuous symmetry.
In fact the infinite volume limit of $P_L(q)$
(see eq. (\ref{12})) is a Bessel
function in contrast with the cosine function which appears in the Ising
case. Taking into account this result and writing the universal function
$f$ in the broken phase as a function of the chiral condensate

\begin{eqnarray}
{f(qL^{-\beta\over{\nu}},{\xi\over{L}})} =
{F(qL^{-\beta\over{\nu}},{{c_{0}^{-\nu\over{\beta}}}\over{L}})}.
\label{16}
\end{eqnarray}

\noindent
the following expression for the universal function $F(z,z')$ holds

\begin{eqnarray}
F(z,z') = {J_{0}({z \over z'^{{\beta \over \nu}}}) + ............},
\label{17}
\end{eqnarray}
\noindent
where $J_0$ is the Bessel function and the dots in (\ref{17})
stand for size
dependent terms which vanish in the infinite volume limit. There are
many ways to parameterize the first corrections in (\ref{17}),
the simplest one
being a term proportional to $q^2$ as follows from the fact that the
universal function $F$ and the Bessel function $J_0$ differ in an analytical
function of $q^2$, which vanish at the origin in momentum space. A detailed
analysis in this direction for some specific models will be reported in a
separate paper. There are however some interesting features concerning
finite volume effects which can be derived from the results reported in
Fig. 2, and which we would like to point out.

Our analysis of the Schwinger model
shows that, as expected, finite volume effects are more or less relevant
depending on the quantity computed. By looking at Fig. 2 we can understand
why finite size effects are relatively large when computing moments of
$P(c)$, like equation (\ref{13}). In fact
in this case all the zeroes of $P(q)$,
or equivalently all the eigenvalues of the fermionic matrix, give contribution
to the moments of $P(c)$ and our results show clearly that higher order
zeroes of $P(q)$ suffer from stronger finite size effects.
What we find certainly surprising in the results
of Fig. 2 is the fact that an accurate estimation of the chiral condensate
in the Schwinger model can be obtained from rather small lattices, by
analyzing the first zeroes of $P(q)$. They show an asymptotic scaling
behavior which follows from the fact that the first eigenvalues of the
dominant gauge field configurations scale with the inverse lattice volume,
the window for this scaling increasing with the lattice size. These results
should stimulate people working in this field to apply this formalism to
more interesting physical systems, like $QCD$.

We thank CICYT (Spain) for partial financial support to this work.

\endpage

\endpage
\vskip 1 truecm
\leftline{\bf Figure captions}
\vskip 1 truecm
{\bf Figure 1.} Standard form of $P(c)$ (eq. (1.11)) in the broken phase.

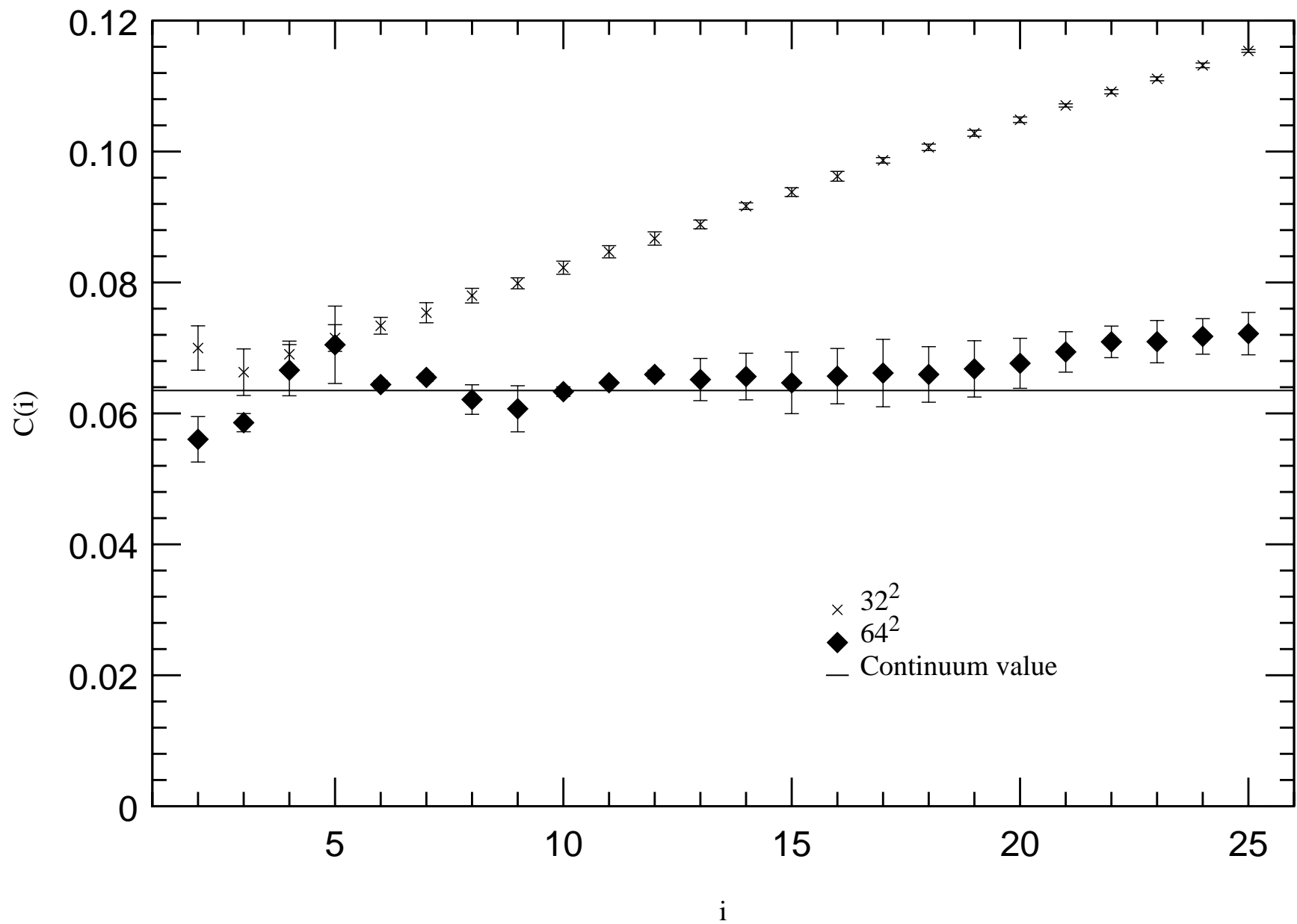
{\bf Figure 2.} Chiral order parameter in the compact
Schwinger model in $32^2$ and $64^2$ lattices at $\beta=6.344$. The solid line
corresponds to its continuum analytical value.

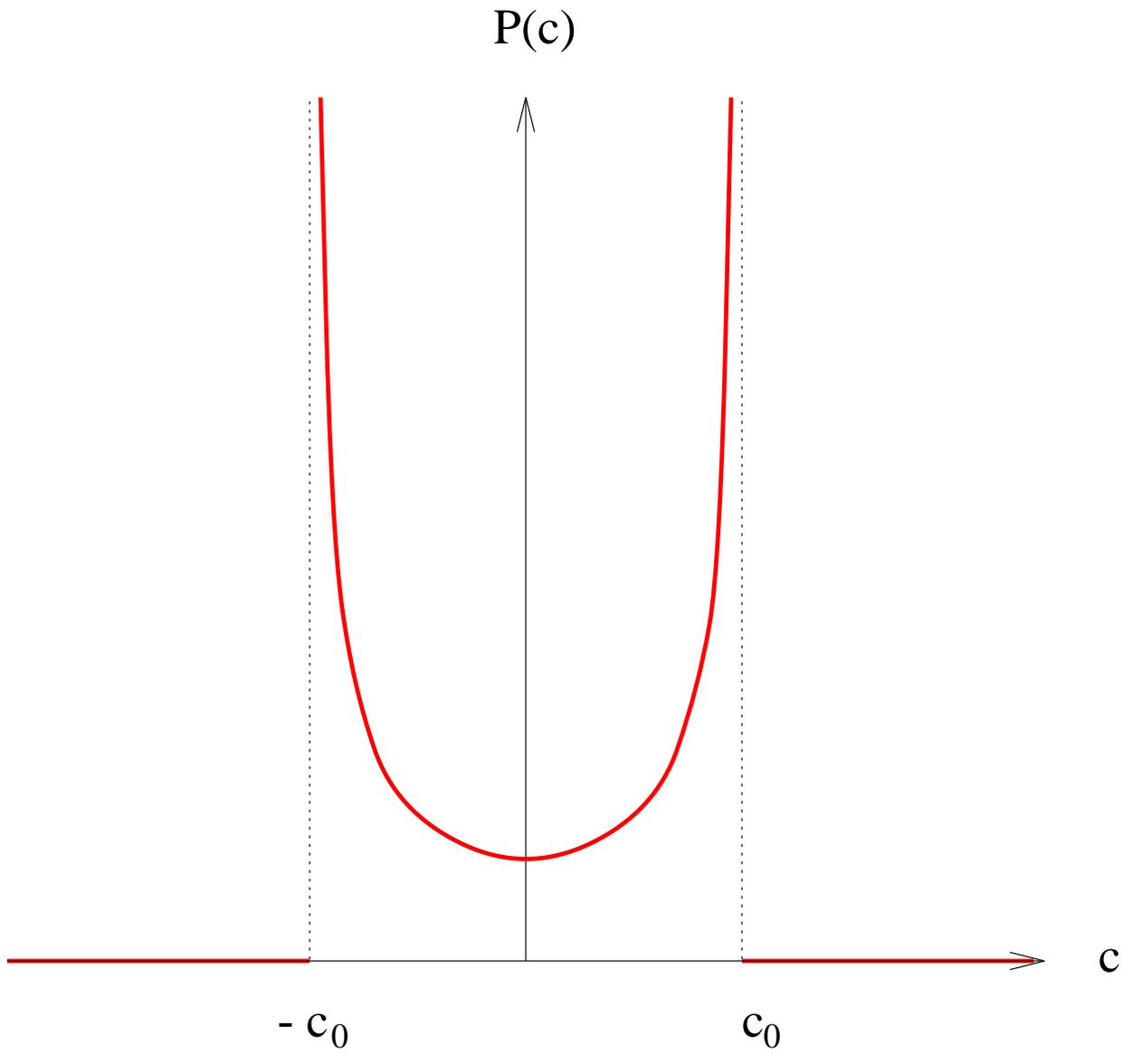
\end{document}